\documentclass[onecolumn,draftclsnofoot]{IEEEtran}

\usepackage{times,amsmath,epsfig,latexsym,amssymb,psfrag,booktabs,extpfeil,fancyhdr}
\begin{document}
\linespread{1.5}
\title{Link Adaptation with Untrusted Relay Assignment: Design and Performance Analysis}
\author{\IEEEauthorblockN{Hamid Khodakarami, Farshad Lahouti\\}
\IEEEauthorblockA{Center for Wireless Multimedia Communications\\
School of Electrical and Computer Engineering, University of Tehran\\
Email: khodakarami@ut.ac.ir, lahouti@ut.ac.ir}\thanks{A preliminary report on this work has been presented at the IEEE International Conference on Telecommunications, Cyprus, May 2011 \cite{UntrustedRelayConferenceICT:Khodakarami}.}}
\maketitle
\linespread{1.9}
\begin{abstract}
In this paper, a link adaptation and untrusted relay assignment (LAURA) framework for efficient and reliable wireless cooperative communications with physical layer security is proposed. Using sharp channel codes in different transmission modes, reliability for the destination and security in the presence of untrusted relays (low probability of interception) are provided through rate and power allocation. Within this framework, several schemes are designed for highly spectrally efficient link adaptation and relay selection, which involve different levels of complexity and channel state information requirement. Analytical and simulation performance evaluation of the proposed LAURA schemes are provided, which demonstrates the effectiveness of the presented designs. The results indicate that power adaptation at the source plays a critical role in spectral efficiency performance. Also, it is shown that relay selection based on the signal to noise ratio of the source to relays channels provides an interesting balance of performance and complexity within the proposed LAURA framework.  
\end{abstract}
\setlength{\headheight}{24pt}
\thispagestyle{fancy}
\lhead{\textit{This manuscript has been submitted to IEEE for possible publication as a journal paper. It is therefore subject to IEEE copyright regulations.}}
\renewcommand{\headrulewidth}{0.0pt}
\IEEEpeerreviewmaketitle

\providecommand{\abs}[1]{\lvert#1\rvert}
\newtheorem{remark0}{Remark}

\section{Introduction}
\IEEEPARstart{D}{iversity} and link adaptation are two enabling techniques to facilitate high performance communications over wireless fading channels. In cooperation for diversity, the relays are to assist a reliable data transmission from the source to the destination. One major challenge hindering the practical adoption of cooperative wireless communications is the security as the relaying nodes may in fact be able to eavesdrop on the source destination communications. In this paper, wireless link adaptation solutions are proposed to facilitate both reliability and physical layer security for cooperative communications in presence of untrusted relays.


The cooperation of the source with a relay node may include a so-called \emph{service level} trust, i.e., the relay node indeed performs its expected function as a relay in the network. However, this cooperation may not necessarily include a \emph{data level} trust, i.e., the relay may not be supposed to extract (decode) useful information from the source destination communication. 
The information theoretic aspect of this problem is investigated in \cite{CodingForRelayChannelConfMess:Oohama}, \cite{CooperetiveUntrustedRelay:Yener}.  
Specifically, an upper bound for the achievable secrecy rate in this setting is presented in  \cite{CodingForRelayChannelConfMess:Oohama}.

Link adaptation by rate and power control could highly improve the performance of cooperative communications over time-varying channels. In \cite{OnCapacityFadingcooperetiveAdaptTransmission:Nguyen}, the capacity of  adaptive transmission over cooperative fading channel is considered for amplify-and-forward relaying, where three different adaptive techniques are investigated. In \cite{PerformanceAnalIncre:Alouini}, the performance of cooperative communications with relay selection and un-coded adaptive modulation is investigated. An scheme for joint power and bandwidth allocation and relay selection in a user cooperative network is proposed in \cite{JointRelayResource:WeiYu}, which considers optimizing a utility function of user traffic demands in a slow but frequency selective fading channel. In \cite{SingleMultipleRelaySelection:JafarKhani}, single and multiple relay cooperation with amplify-and-forward (AF) relaying is considered. For reduced complexity, suboptimal multiple relay selection schemes are proposed and shown to achieve full diversity. In \cite{JointAMCARQCoop:Mardani}, a cross-layer approach to optimize the spectral efficiency of the relay channel employing adaptive modulation and coding in conjunction with cooperative automatic repeat request is proposed.

A practical code design approach to physical layer security based on capacity achieving low density parity check (LDPC) codes is introduced in \cite{SecureLDPCBEC:McLaughlin}. The setting is a wiretap binary erasure channel involving a source, a destination and an eavesdropper. In fact, constructive approaches for (imperfect) physical layer security may be set up to ensure a reliable source destination communication, while maintaining a high probability of error for the eavesdropper. To this end, design of sharp punctured LDPC codes for which the bit error rate (BER) curve falls sharply from high BERs to very low BERs (steep waterfall region), is considered in \cite{LDPCforAWGN:Mclaughlin}. Yet, adaptive transmission may be exploited to enhance the system performance over time-varying channels both in terms of security and reliability \cite{IETLinkadaptation:khodakarami}.

In this paper, a framework for cooperative communications through link adaptation with untrusted relay assignment (LAURA) is proposed. The purpose is to utilize the cooperation of arbitrary number of relays for reliable communications, while ensuring they cannot decode useful information from their relayed signal. Different LAURA schemes are set up based on network CSI that is already necessary for quality of service (QoS) provisioning over wireless fading channels. The resource degrees of freedom, i.e, transmission power, transmission rate, and cooperating relays, are exploited in an optimized manner to address the QoS reliability and security requirements of the cooperative communication networks dealing with untrusted relays. Specifically, several  power adaptation and relay selection strategies are proposed for LAURA that are designed for high spectral efficiency communication based on discrete rate adaptation with sharp channel codes and different levels of complexity an
 d CSI requirements. Analytical and simulation performance evaluation of the proposed LAURA schemes are provided, which demonstrates the effectiveness of the presented designs. The results indicate that the LAURA design with source (only) power adaptation and relay selection based on (only) source relay channels CSI provide an interesting balance of performance and complexity within the proposed LAURA framework. 

The paper is organized as follows. Section \ref{Sec_SystemModelSection} presents the system model and describes the problem under consideration. In Section \ref{Sec_Original_Prob_and_Solve}, the link adaptation and relay selection problem is introduced and its exact solution is presented. Section \ref{Sec_Constant_Power_Strategies} explores the scenarios where source and relays transmit with constant power. In Section \ref{Sec_Relay_Selection_Strategies}, suboptimal relay selection strategies are investigated. Section \ref{Sec_Performance_Evaluation} presents the simulation and theoretical results for the proposed LAURA schemes. Section \ref{ConclusionSection} concludes the paper.
\section {System Model and Problem Statement}\label{Sec_SystemModelSection}
\newcommand{\ud}{\mathrm{d}}
\begin{figure}[!t]
\centering
\includegraphics[width=3in]{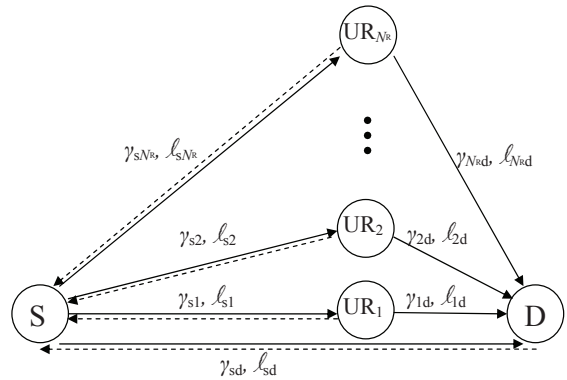}
\caption{Network topology; $\gamma$ and $\ell$ indicate the channel SNR and the distance.}
\label{Network}
\end{figure}
A wireless communication system with one source node (S), one destination node (D) and a set of $N_{\text{R}}$ available relay nodes, $\left\{\mathrm{UR}_i|,i\in \mathcal{M}_{\text{R}}=\{1,2,...,N_{\text{R}}\}\right\}$, is considered (Fig. \ref{Network}). There are $N_{\text{C}}$ cooperating relays, denoted by the set $\mathcal{M}_{\text{C}}$, that are selected from the set of available relays $\mathcal{M}_{\text{R}}$.  The cooperation protocol is AF. The rest of the relays in the set $\mathcal{M}_{\text{R}}$-$\mathcal{M}_{\text{C}}$ are referred to as non-cooperating relays. 
The relays are assumed trusted at the service level and untrusted at the data level. Service level trust entails following the AF protocol as expected. This involves for relays to feedback true CSI, remain inactive if not selected for cooperation, and upon selection for cooperation, adapting their power according to source schedule, and forwarding the amplified version of received signal without modification. Since the relays are data level untrusted, the source imposes security constraints on relays. This is to prevent all relay nodes from extracting useful information from their received signal. 

Total network security calls for an array of technologies and steps involving different layers of communications protocol stack, and including hardware, process and physical security. For instance, service level security could be facilitated in part by hardware and physical security to avoid hardware tampering. In this work, we focus on data level security that can be provided by link adaptation and relay assignment in wireless AF relay networks. Needless to say, this will serve as one (added) layer of security and in general does not make other security mechanisms evadable.

In the first phase of the transmission, the S node transmits signal $x$ to the D node and relays. The received signal at the D node and at the $\mathrm{UR}_i$ node are respectively
\begin{align}
y_{\text{sd}} &= h_{\text{sd}}\,x+w_{\text{d}}\\
y_{\text{s}i} &= h_{\text{s}i}\,x+w_{\text{s}i},
\end{align}
where $h_{\text{sd}}$, and $h_{\text{s}i}$ denote the Rayleigh fading coefficients between S and D nodes, and S and $\mathrm{UR}_i$ nodes, respectively. The noise is denoted at the  $\mathrm{UR}_i$ node as $w_{\text{s}i}$ and at the D node as $w_{\text{d}}$. The $N_{\text{C}}$ cooperating relays amplify the received signal and transmit it to the D node in the second phase of cooperation. The received signal at the D node from the $\mathrm{UR}_i$ node is
\begin{equation}
y_{i\text{d}} = G_i\, h_{i\text{d}}\, y_{\text{s}i} + w_{\text{d}}',
\end{equation}
where $G_i$ is the $\mathrm{UR}_i$ node amplifier gain and $h_{i\text{d}}$ and $w'_{\text{d}}$ are Rayleigh fading coefficients from $\mathrm{UR}_i$ node to the D node and noise at the D node, respectively.

The S-D, S-$\mathrm{UR}_i$ and $\mathrm{UR}_i$-D channels, for $1 \leq i \leq N_{\text{R}}$, are independent Rayleigh fading channels with SNRs of $\gamma_{\text{sd}}$, $\gamma_{\text{s}i}$ and $\gamma_{i\text{d}}$, respectively. The SNRs are exponentially distributed with parameters  $\frac{1}{\bar\gamma_{\text{sd}}}$, $\frac{1}{\bar\gamma_{\text{s}i}}$ and $\frac{1}{\bar\gamma_{i\text{d}}}$ and, probability density functions of $f_{\text{sd}}(\gamma_{\text{sd}})$, $f_{\text{s}i}(\gamma_{\text{s}i})$ and $f_{i\text{d}}(\gamma_{i\text{d}})$, respectively. With full CSI assumption, the instantaneous SNRs of the three channels are assumed known at the onset of each frame interval at the S node. Assuming maximum ratio combining (MRC) of signals received from S node and $N_{\text{C}}$ cooperating relay nodes at the D node, the equivalent SNR of AF relaying protocol at the D node is \cite{CooperativeCommNetworking:RayLiu}
\begin{equation}\label{Eq_MRC_All_Relays}
\gamma_{\text{eq}} = \frac{S_{\text{s}}}{S}\gamma_{\text{sd}} + \sum_{i\in \mathcal{M}_{\text{C}}} \gamma_{i},
\end{equation}
where
\begin{equation}\label{Eq_SNR_DualHop}
\gamma_{i}=\frac{\frac{S_{\text{s}}}{S}\gamma_{\text{s}i}\,\frac{S_i}{S}\gamma_{i\text{d}}}{\frac{S_{\text{s}}}{S}\gamma_{\text{s}i}+\frac{S_i}{S}\gamma_{i\text{d}}+1},
\end{equation}
where $S_{\text{s}}$ and $\boldsymbol{S}=\{S_i|i\in \mathcal{M}_{\text{C}}\}$ are transmission power of source and the set of transmission powers of relays, respectively. Here, $S$ normalizes the transmission power to the case where no power adaptation is employed. For the sake of tractability of theoretical performance analyses in some cases, we may use an upper bound on $\gamma_i$ as follows  \cite{OnCapacityFadingcooperetiveAdaptTransmission:Nguyen}-\cite{PerformanceAnalIncre:Alouini}
\begin{equation}\label{Eq_MRC_UB}
\gamma_{i,\text{u}}=\min(\frac{S_{\text{s}}}{S}\gamma_{\text{s}i},\frac{S_i}{S}\gamma_{i\text{d}}). 
\end{equation}
This yields the following upper bound on the equivalent SNR at the destination $\gamma_{\text{eq,u}}=\frac{S_{\text{s}}}{S}\gamma_{\text{sd}} + \sum_{i\in \mathcal{M}_{\text{C}}} \gamma_{i,u}$. 

We use a set of $N$ transmission modes (TM) each corresponding to a combination of modulation and coding. These TMs provide transmission rates of $R_1, R_2, ... , R_N$ bits per symbol and we assume $R_n >R_{n-1}$. The LAURA schemes proposed in the sequel may be set up based on any given set of coding and modulation pairs, as for their design we assume closed-form expressions for the performance of TMs. Indeed, we can express the instantaneous BER of TM $n$ as an approximated function of received SNR, $\gamma$, through curve fitting by
\begin{equation*}\label{BERApproximation}
\mathrm{IBER}_n(\gamma) = \left\{
\begin{array}{lr}
0.5\, e^{-p_n\gamma^{q_n}}\hspace{10mm} & \text{if } \gamma < \gamma_{lh}^n\\
\dfrac{a_n}{\left({1+e^{c_n(\gamma-b_n)}}\right)^{k_n}}  & \text{if
} \gamma \geq \gamma_{lh}^n
\end{array} \right.
\end{equation*}
where $p_n$, $q_n$, $a_n$, $b_n$, $c_n$ and $k_n$ are the approximation fitting parameters for a given modulation and coding pair. The value of $\gamma_{lh}^n$ is determined by the intersection of the two parts of the approximation. The expression for the second part is a modification of what is proposed in \cite{PerformLDPC:Mackay}, and the first part is devised here for better performance modeling of TMs in the low SNR regime. The inverse of $\mathrm{IBER}_n(\gamma)$ describing received SNR as a function of BER is given by
\begin{equation}\label{SNRapproximation}\
\Gamma_n(P_e)\! =\!\left\{
\begin{array}{lr}
\!\!\!\!\left(\ln(\dfrac{0.5}{ P_e})\Big/p_n\right)^{\frac{1}{q_n}}\hspace{5mm}&\!\!\!\text{if } P_e >\mathrm{IBER}_n(\gamma_{lh}^n)\\
\!\!\!\!\frac{1}{c_n}\ln\Biggl[\left(\dfrac{a_n}{P_e}\right)^{1/k_n}\!\!\!\!-1\Biggr] +b_n&
\!\!\!\text{if }P_e\leq \mathrm{IBER}_n(\gamma_{lh}^n).
\end{array}\right.
\end{equation}

The objective is to maximize the spectral efficiency of the system while providing both security against eavesdropping of data level untrusted relays and reliable communications for the destination. These two requirements are expressed as follows
\begin{align}\label{Eq_Security_Reliability_Const}
\mathrm{C1.}\quad & \mathrm{BER}_i\geq \mathrm{BER}_{\text{tgt}}^{\text{r}} \,\,\,\text{for}\, 1\leq i\leq N_{\text{R}} \nonumber\\
\mathrm{C2.}\quad &\mathrm{BER}_{\text{d}}\leq\mathrm{BER}_{\text{tgt}}^{\text{d}},
\end{align}  
where $\mathrm{BER}_i$ is the BER at the $\mathrm{UR}_i$ node and $\mathrm{BER}_{\text{d}}$ is the BER at the D node applying MRC. In \eqref{Eq_Security_Reliability_Const}, $\mathrm{BER}_{\text{tgt}}^{\text{r}}$ is a target lower limit for $\mathrm{BER}_i$ to ensure security against eavesdropping of relays, and $\mathrm{BER}_{\text{tgt}}^{\text{d}}$ is a target upper limit for BER at the destination to ensure reliable communications.
The average spectral efficiency is expressed as
\begin{align}\label{Eq_Spectral_Eff_General}
\eta &= \sum_{n=1}^N \frac{R_n}{2}\, \mathrm{Pr}(\mathrm{TM} = n),
\end{align}
where $\mathrm{Pr}(\mathrm{TM} = n)=P_n$ is the probability of selecting TM number $n$ for transmission. The factor $1/2$ multiplied by $R_n$ is due to the half-duplex cooperative transmission. The equation \eqref{Eq_Spectral_Eff_General} indicates that to compute the average spectral efficiency of the system, we only need to derive an expression for $P_n$.
\section{Link Adaptation and Untrusted Relay Assignment for Cooperative Communications}\label{Sec_Original_Prob_and_Solve}
The general optimization problem for LAURA with power adaptation is expressed as follows
\begin{align}\label{Eq_Opt_Prob_General}
&\max_{\mathcal{M}_{\text{C}}}\max_{n\in\{1,2,...,N\}}\max_{S_{\text{s}},\boldsymbol{S}}\quad R\\
\mathrm{s.t.} \qquad
&\mathrm{C1.}\quad  \mathrm{BER}_{j}\geq \mathrm{BER}_{\text{tgt}}^{\text{r}}\, \text{for}\, 1\leq j\leq N_{\text{R}} \nonumber\\
&\mathrm{C2.}\quad \mathrm{BER}_{\text{d}}\leq\mathrm{BER}_{\text{tgt}}^{\text{d}} \nonumber\\
&\mathrm{C3.}\quad  S_{\text{s}}+ \sum_{i\in \mathcal{M}_{\text{C}}}S_i \leq S_{\text{tot}},\nonumber
\end{align}
where $R$ is the transmission rate in a given frame in bits per symbol or equivalently the spectral efficiency. Note that the solution to this problem provides the instantaneous (per frame) power allocation scheme in the two transmission phases. The constraint $\mathrm{C3}$ in \eqref{Eq_Opt_Prob_General} expresses a network sum power constraint, where the total transmission of source and cooperating relays are limited. This is of course justified given that the relays are service level trusted (Section \ref{Sec_SystemModelSection}), and in line with many articles in literature, e.g., \cite{Decode:Luo,Bandwidth:Maric}, allows us to better understand the potential benefit of (additional) relays given a certain network power budget. As we shall discuss in Remark \ref{Remark_Separate_Power_Const}, the case with separate power constraints for the source and each of the relays will simply be a special case. An instance of the problem with $N_{\text{C}}=1$ is considered in \cite{UntrustedRelayConferenceICT:Khodakarami}, where a single relay is selected for cooperation out of the $N_{\text{R}}$ available relays. 
In the following, the exact solution to the problem \eqref{Eq_Opt_Prob_General} is presented and in the next section we explore some other possible solutions. A list of the proposed LAURA schemes and a summary of their characteristics are provided in Table \ref{CSIrequirement}.
\linespread{1.5}
\begin{table}[!t]
\centering 
\caption{CSI and power optimization requirements of different proposed schemes. AAR, ACR and ASR denote all available relays, all cooperating relays and all $N_{\text{C}}$ member subsets of relays, respectively.}\label{CSIrequirement}
\begin{tabular}{|c|c|c|c|}
\hline
\textbf{Scheme}&\textbf{Relay Selection}&\textbf{CSI requirements}&\textbf{Power Optimization}\\
\hline
\vspace{-4mm}  & &\\ \hline
LAURA1 & Optimal & $\gamma_{\text{s}i}$ (for AAR),  $\gamma_{i\text{d}}$ (for AAR),  $\gamma_{\text{sd}}$ & Source \& ASR \\
LAURA2& Based on S-$\mathrm{UR}_i$ CSI&$\gamma_{\text{s}i}$  (for AAR),  $\gamma_{i\text{d}}$ (for ACR),  $\gamma_{\text{sd}}$ & Source \& ACR \\
LAURA1-CPR& Optimal & $\gamma_{\text{s}i}$ (for AAR),  [$\gamma_{\text{eq}}$ (for ASR)] or [$\gamma_{\text{sd}}$ and  $\gamma_{i\text{d}}$ (for AAR)]& Source\\
LAURA2-CPR& Based on S-$\mathrm{UR}_i$ CSI & $\gamma_{\text{s}i}$ (for AAR),  [$\gamma_{\text{eq}}$ (for ACR)] or [$\gamma_{\text{sd}}$ and  $\gamma_{i\text{d}}$ (for ACR)]& Source\\
LAURA3-CPR& Based on Average S-$\mathrm{UR}_i$ SNR &$\gamma_{\text{s}i}$ (for AAR),   [$\gamma_{\text{eq}}$ (for ACR)]  or [$\gamma_{\text{sd}}$  and $\gamma_{i\text{d}}$ (for ACR)]& Source\\
LAURA1-CP& Optimal &$\gamma_{\text{s}i}$ (for AAR),   [$\gamma_{\text{eq}}$ (for ASR)] or [$\gamma_{\text{sd}}$ and $\gamma_{i\text{d}}$ (for AAR)] & N/A \\
\hline
\end{tabular}
\end{table}
\linespread{1.9}
\subsection{Exact Solution}
The problem \eqref{Eq_Opt_Prob_General} can be solved exactly and the resulting solution which serves as an upper-bound on performance is referred to as LAURA1 in the sequel. In this case, the CSI of S-D, S-$\mathrm{UR}_i$ and $\mathrm{UR}_i$-D are required at the S node at the beginning of each transmission interval. The following proposition enables the proposed solution.
\newtheorem{proposition1}{Proposition}
\begin{proposition1}
While $\mathrm{C2}$ is to be satisfied in \eqref{Eq_Opt_Prob_General}, maximizing $R_n$  is equivalent to maximizing $\gamma_{\text{eq}}$.
\end{proposition1}
\begin{proof}
Consider $\gamma_{\text{eq}}^*$ as the optimized $\gamma_{\text{eq}}$ constrained to $\mathrm{C2}$. Suppose that $\gamma_{\text{eq}}'<\gamma_{\text{eq}}^*$. 
Now, if \linebreak $n'=\mathrm{arg}\max_n\{\mathrm{IBER}_n(\gamma_{\text{eq}}')\leq \mathrm{BER}_{\text{tgt}}^{\text{d}}\}$, i.e., the maximum value of $n$ given that $\mathrm{IBER}_n(\gamma'_{\text{eq}})$ is less than or equal to
$\mathrm{BER}^{\text{d}}_{\text{tgt}}$, and $n^*=\mathrm{arg}\max_n\{\mathrm{IBER}_n(\gamma_{\text{eq}}^*)\leq \mathrm{BER}_{\text{tgt}}^{\text{d}}\}$, then $R_{n^*}\geq R_{n'}$. 
\end{proof}

Using Proposition 1, for a given set $\mathcal{M}_{\text{C}}$ and a given TM $n$, the following design subproblem is considered.  
\begin{align}\label{Eq_Opt_Prob_Gamma_eq_General}
&\max_{S_{\text{s}},\boldsymbol{S}}\quad \gamma_{\text{eq}}\\
\mathrm{s.t.} \qquad
&\mathrm{C1.}\quad  S_{\text{s}}\leq S \frac{\Gamma_n^{\text{r}}}{\gamma_{\text{s}\tilde i}} \nonumber\\
&\mathrm{C3.}\quad  S_{\text{s}}+ \sum_{i\in \mathcal{M}_{\text{C}}}S_i \leq S_{\text{tot}}\nonumber,
\end{align}
where $\tilde i=\mathrm{arg}\max_i \gamma_{\text{s}i}$ and based on \eqref{SNRapproximation}, $\mathrm{\Gamma}_n^r \triangleq\Gamma_n(  \mathrm{BER}_{\text{tgt}}^{\text{r}})$. Setting $\tilde i$ as such ensures security against eavesdropping of any of the relays. As presented below, the solution to this problem provides optimized instantaneous power allocation at the source and the set of cooperating relays under consideration. Next, we use the results within Algorithm 1, which yields the optimum TM $n$ and choice of cooperating relays $\mathcal{M}_{\text{C}}$ in presence of the constraint $\mathrm{C2}$ in \eqref{Eq_Opt_Prob_General}. In fact, Algorithm 1 identifies the largest TM $n$ (rate) for which the set of cooperating relays leading to maximum $\gamma_{\text{eq}}$ in subproblem \eqref{Eq_Opt_Prob_Gamma_eq_General} satisfies $\mathrm{C2}$ in \eqref{Eq_Opt_Prob_General}.

It can be easily shown that $\gamma_{\text{eq}}$ is a concave function of $(S_{\text{s}},\boldsymbol{S})$. 
In other words, the next three conditions for $\gamma_{\text{eq}}$ hold ‎\cite[Appendix 1]{OpereatoinalResearchBook:Hillier}: ‎$‎\partial^2\gamma_{\text{eq}}‎ / ‎\partial‎ ‎S_{i}^2‎\leq‎ ‎0$, $\partial^2\gamma_{\text{eq}}‎/ ‎ \partial‎ ‎S_{\text{s}}^2‎\leq‎ ‎0‎$‎‎ and ‎$(‎\partial^2\gamma_{\text{eq}} ‎/ ‎\partial‎ ‎S_{i}^2)(\partial^2\gamma_{\text{eq}}‎/‎\partial‎ ‎S_{\text{s}}^2‎‎)-‎[‎‎(‎\partial^2\gamma_{\text{eq}}‎/‎\partial‎ ‎S_{\text{s}}\partial‎ ‎S_i)‎]‎^2‎‎‎‎\leq‎ ‎0‎‎$‎‎, for $i\in \mathcal{M}_{\text{R}}$. Thus, the KKT condition gives the optimal solution to problem \eqref{Eq_Opt_Prob_Gamma_eq_General}. 
In order to simplify the solution, we first solve \eqref{Eq_Opt_Prob_Gamma_eq_General} without $\mathrm{C1}$. The governing Lagrangian for this problem temporarily ignoring $\mathrm{C1}$ is
\begin{equation}
\mathcal{L} = \gamma_{\text{eq}}+ \lambda_1 \left( S_{\text{s}}+ \sum_{i\in \mathcal{M}_{\text{C}}}S_i - S_{\text{tot}}\right).
\end{equation} 
The optimal $S_{\text{s}}$ and $S_i$ should satisfy 
\begin{equation}
\frac{\partial\mathcal{L}}{\partial S_{\text{s}}} =0 \,\,\text{and}\,\, \frac{\partial\mathcal{L}}{\partial S_{i}} =0\quad \text{for}\, i\in \mathcal{M}_{\text{C}}.
\end{equation} 
Then, 
\begin{equation}
\frac{\partial\mathcal{L}}{\partial S_{i}} =  \frac{\dfrac{\gamma_{\text{s}i}^2\gamma_{i\text{d}}}{S^3}S_i^2+\dfrac{\gamma_{\text{s}i}\gamma_{i\text{d}}}{S^2}S_i}{\left(\frac{S_{\text{s}}}{S}\gamma_{\text{s}i}+\frac{S_i}{S}\gamma_{i\text{d}}+1\right)^2}+\lambda_1=0,
\end{equation}
that yields
\begin{align}\label{Eq_S_i_function_of_Lambda_S_s_General}
S_i= \frac{S}{\gamma_{i\text{d}}}\left[\frac{1}{\nu}\left(\frac{S_{\text{s}}^2}{S^3}\gamma_{\text{s}i}^2 \gamma_{i\text{d}}+\frac{S_{\text{s}}}{S^2}\gamma_{\text{s}i}\gamma_{i\text{d}}\right)^{1/2}-\frac{S_{\text{s}}}{S}\gamma_{\text{s}i}-1\right]^+,
\end{align}
where $\nu=\sqrt{-\lambda_1}$ and $[x]^+$ denotes $\max(x,0)$.
Also,
\begin{equation}\label{Eq_d_L_d_S_s_General}
\frac{\partial\mathcal{L}}{\partial S_{\text{s}}}=\frac{\gamma_{\text{sd}}}{S}+ \sum_{i\in \mathcal{M}_{\text{C}}} \frac{\dfrac{\gamma_{\text{s}i}\gamma_{i\text{d}}^2}{S^3}S_{\text{s}}^2+\dfrac{\gamma_{\text{s}i}\gamma_{i\text{d}}}{S^2}S_{\text{s}}}{\left(\frac{S_{\text{s}}}{S}\gamma_{\text{s}i}+\frac{S_i}{S}\gamma_{i\text{d}}+1\right)^2}+\lambda_1=0.
\end{equation}
Substitution of \eqref{Eq_S_i_function_of_Lambda_S_s_General} in $\mathrm{C3}$ of \eqref{Eq_Opt_Prob_Gamma_eq_General} and \eqref{Eq_d_L_d_S_s_General} gives a set of two equations. Numerically solving this set of equations yields the optimal $S_i$ for $i\in \mathcal{M}_{\text{C}}$  and also  $S_{\text{s}}$ for \eqref{Eq_Opt_Prob_Gamma_eq_General} considering only $\mathrm{C3}$. Now we check whether the solution satisfies  $\mathrm{C1}$ in \eqref{Eq_Opt_Prob_Gamma_eq_General}. If the condition is satisfied, then the solution is the optimal one, else according to $\mathrm{C1}$ we set  $ S_{\text{s}}= S \frac{\Gamma_n^{\text{r}}}{\gamma_{\text{s}\tilde i}}$ in \eqref{Eq_S_i_function_of_Lambda_S_s_General}. Next, using the result in $\mathrm{C3}$  of \eqref{Eq_Opt_Prob_Gamma_eq_General} yields the optimum $\boldsymbol{S}$ by quantifying the new value of $\nu$.
Using the presented solution to \eqref{Eq_Opt_Prob_Gamma_eq_General}, as described, the Algorithm 1 formulates the LAURA1 scheme or the exact solution to design problem \eqref{Eq_Opt_Prob_General}. In this algorithm, we use the optimal power allocation for source and relays for every $N_{\text{C}}$ member subset of the available relays as possible cooperating ones, and choose the best subset according to the equivalent SNR it provides. Based on this, we find the highest possible transmission rate such that both security and reliability constraints are satisfied.  
\vspace{2mm}\\
\begin{tabular}{l}
\toprule[0.5mm]
Algorithm 1: Exact Solution for LAURA\\
\midrule
1) Select $n=N$.\\
2) If $n=0$ then go to outage mode and exit.\\
3) Build the set $\mathcal{M}_{\text{A}}$ comprised of all $N_{\text{C}}$ member subsets of $\mathcal{M}_{\text{R}}$ and index its members as $\mathcal{M}_{\text{C},l}$,\\ $l \in \{1,2,...,\frac{N_{\text{R}}!}{(N_{\text{R}}-N_{\text{C}})!N_{\text{C}}!}\}$.\\
4) For $l=1$ to $\frac{N_{\text{R}}!}{(N_{\text{R}}-N_{\text{C}})!N_{\text{C}}!}$\\
\hspace{8mm}4-1) Set $\mathcal{M}_{\text{C}}=\mathcal{M}_{\text{C},l}$\\
\hspace{8mm}4-2) Solve \eqref{Eq_Opt_Prob_Gamma_eq_General} and obtain $S_{\text{s}}^{(l)}$ and $S_i^{(l)}$, $i\in \mathcal{M}_{\text{C}}$\\ 
\hspace{8mm}4-3) Calculate $\gamma_{\text{eq}}^{(l)}$.\\
5) Select the set of cooperating relays by $l^* = \mathrm{arg}\max_l \gamma_{\text{eq}}^{(l)}$ and set $\gamma_{\text{eq}}=\gamma_{\text{eq}}^{(l^*)}$.\\
6) If $\mathrm{C2}$ in \eqref{Eq_Opt_Prob_General} is satisfied, set $\mathrm{TM}=n$ and exit; else $n=n-1$ and go to step $2$.
\vspace{1mm}\\
\bottomrule[0.5mm]
\end{tabular}
\vspace{2mm}

\subsection{Analytical Results for $N_{\text{C}}=1$}
In case only one relay is to be selected for cooperation, i.e, $N_{\text{C}}=1$, closed form solutions for transmission powers of source and relay can be obtained. This in turn allows for an analytical performance evaluation in this case. For $N_{\text{C}}>1$, we resort to numerical results for performance evaluation in Section  \ref{Sec_Performance_Evaluation}. 

Considering the design subproblem in \eqref{Eq_Opt_Prob_Gamma_eq_General} for $N_{\text{C}}=1$ and using $\mathrm{UR}_i$ node for cooperation, we have $S_i=S_{\text{tot}}-S_{\text{s}}$ and \eqref{Eq_d_L_d_S_s_General} yields $ S_{\text{s},n,i}^{\text{*}}=\frac{-\theta_i-\sqrt{\theta_i^2-4\mu_i\rho_i}}{2\mu_i}$, where 
\begin{align*}
\mu_i&\! =\! \gamma_{i\text{d}}^2(\gamma_{\text{s}i} + \gamma_{\text{sd}}) +\gamma_{\text{s}i}^2(\gamma_{\text{sd}}- \gamma_{i\text{d}}) - 2\gamma_{\text{sd}}\gamma_{i\text{d}}\gamma_{\text{s}i}\\
\theta_i&\! =\!2[\gamma_{\text{s}i}(\gamma_{\text{sd}}-\! \gamma_{i\text{d}}+ \!\gamma_{i\text{d}}\gamma_{\text{sd}}S_{\text{tot}})-\gamma_{i\text{d}}^2 S_{\text{tot}}( \! \gamma_{\text{s}i} +\! \gamma_{\text{sd}})\!-\! \gamma_{i\text{d}}\gamma_{\text{sd}}]\\
\rho_i& \!=\! \gamma_{\text{sd}} + \gamma_{i\text{d}}^2(\gamma_{\text{s}i} + \gamma_{\text{sd}})S_{\text{tot}}^2 + (2\gamma_{i\text{d}}\gamma_{\text{sd}} + \gamma_{i\text{d}}\gamma_{\text{s}i})S_{\text{tot}}.
\end{align*}
Then, the optimal transmission power of S node cooperating with $\mathrm{UR}_i$ node is
\begin{equation}\label{S_opt_discreteRate}
S_{\text{s},n,i}^{\text{opt}} =\min(S_{\text{s},n,i}^{*},S_{\text{tot}},S\frac{\Gamma_n^{\text{r}}}{\gamma_{\text{s}\tilde i}}).
\end{equation}
We next use this result in Algorithm 1. Noting the constraint $\mathrm{C2}$ in \eqref{Eq_Opt_Prob_General} for $N_C=1$, we consider the following event, which describes the case when relay $\mathrm{UR}_i$ is selected and TM $n$ satisfies both the reliability and the network security constraints,
\begin{equation}\label{Eq_AP_MR_Reliability_Constraint}
A_n^i : \gamma_{\text{eq}}^{(i)}=\left[ \dfrac{S_{\text{s}}}{S}\gamma_{\text{sd}}+\frac{\frac{S_{\text{s}}}{S}\gamma_{\text{s}i}\,\frac{S_{\text{tot}}-S_{\text{s}}}{S}\gamma_{i\text{d}}}{\frac{S_{\text{s}}}{S}\gamma_{\text{s}i}+\frac{S_{\text{tot}}-S_{\text{s}}}{S}\gamma_{i\text{d}}+1}\right]_{S_{\text{s}} =S_{\text{s},n,i}^{\text{opt}}}\geq \Gamma_n^{\text{d}},
\end{equation}
where $\mathrm{\Gamma}_n^{\text{d}} \triangleq \Gamma_n(\mathrm{BER}_{\text{tgt}}^{\text{d}})$. For the presented transmission strategy in Algorithm 1,  the event that TM number $n$ satisfies both security and reliability requirements is denoted by $A_n$ and its probability is
\begin{align}\label{Eq_Prob_of_Mode_Can_be sel_Adaptive_power2}
\mathrm{Pr} (A_n)&=\mathrm{Pr} \left(\bigcup_{i=1}^{N_{\text{R}}}A_n^i\right)=1-\mathrm{Pr}\left(\bigcap_{i=1}^{N_{\text{R}}}\left(A_n^i\right)^{c} \right)
=1-\mathbf{E}_{\gamma_{\text{sd}}, \gamma_{\text{s}\tilde i}}\left\{\mathrm{Pr}\left(\bigcap_{i=1}^{N_{\text{R}}}\left(A_n^i\right)^{c}|\gamma_{\text{sd}}, \gamma_{\text{s}\tilde i}\right)\right\},\nonumber\\
&=1-\mathbf{E}_{\gamma_{\text{sd}}, \gamma_{\text{s}\tilde i}}\left\{\prod_{i=1}^{N_{\text{R}}}\left[1-\mathrm{Pr}\left(\gamma_{\text{eq}}^{(i)} \geq \Gamma_n^{\text{d}}|\gamma_{\text{sd}},\gamma_{\text{s}\tilde i}\right)\right]\right\},
\end{align}
where $\mathbf{E}_{\gamma_{\text{sd}}, \gamma_{\text{s}\tilde i}}$ denotes the expectation with respect to independent variables  $\gamma_{\text{sd}}$ and $\gamma_{\text{s}\tilde i}$. The last equality in \eqref{Eq_Prob_of_Mode_Can_be sel_Adaptive_power2} results from the independence of $A_n^i$ and $A_n^j$ for $i\neq j$ and given $\gamma_{\text{sd}}$ and  $\gamma_{\text{s} \tilde i}$. The transmission mode $m$ is selected when it is the TM with the largest rate (here equivalently the largest TM) that with optimal power allocation and relay selection can provide both the security and reliability constraints. The probability of selecting TM $m$ may then be computed as follows
\begin{align}\label{Eq_Prob_of_Mode_Ada_Pow_Mult_Relay_SC}
P_m = \mathrm{Pr}\left(\bigcup_{n=m}^N
A_n\right)-\mathrm{Pr}\left(\bigcup_{n=m+1}^N A_n\right)
=\mathbf{E}\left\{\mathbf{I}\left(A_{\tilde n_m}\right)-\mathbf{I}\left(A_{\tilde n_{m+1}}\right)\right\},
\end{align}
where $\mathbf{I}(\cdot)$ is the indicator function that is $\mathbf{I}(E)=1$ if $E$ is true and $\mathbf{I}(E)=0$ if $E$ is false, and 
\begin{align}
\tilde {n}_m(\gamma_{\text{eq}}^{(i)},\gamma_{\text{s}\tilde i})
= \mathrm{arg}\min_{n\in \{m,...,N\}}\left(\gamma_{\text{eq}}^{(i)}\geq\Gamma_n^{\text{d}}\,|\, S_{\text{s}} =S_{\text{s},n,i}^{\text{opt}}\right).
\end{align}
In fact, using $\mathrm{UR}_i$ relay node with optimum power allocations described, $\tilde {n}_m(\gamma_{\text{eq}}^{(i)},\gamma_{\text{s}\tilde i})$ denotes the minimum TM number from the set $\{m,...,N\}$ for which the reliability and security constraints in given realizations of the fading channels are satisfied. Here, in order to facilitate the analysis we introduce an approximation whose effectiveness is shown in Section \ref{Sec_Performance_Evaluation}. We consider 
\begin{equation}\label{n_tilde_Approximation}
\tilde {n}_m (\gamma_{\text{eq}}^{(i)},\gamma_{\text{s}\tilde i})\approx m.
\end{equation}
Then, the probability of TM $m$ is 
\begin{align}\label{Eq_Prob_of_Mode_Adaptive_Power_SC_MR_AP}
P_m=\mathrm{Pr}\left(A_{m}\right)-\mathrm{Pr}\left(A_{m+1}\right).
\end{align}
It can be easily verified that in case there is no security constraint, \eqref{Eq_Prob_of_Mode_Adaptive_Power_SC_MR_AP} holds without any approximation.
Appendix \ref{Joint_PDF_Of_G_si_and_max_G_si} presents the joint PDF of $ \gamma_{\text{s}\tilde i}$ and $\gamma_{\text{s}i}$ that is required to compute $P_m$ in \eqref{Eq_Prob_of_Mode_Adaptive_Power_SC_MR_AP} and hence the average spectral efficiency. The authors also presented the above analysis in \cite{UntrustedRelayConferenceICT:Khodakarami}.

\section{LAURA: Constant Power Strategies}\label{Sec_Constant_Power_Strategies}
The exact solution for \eqref{Eq_Opt_Prob_General} or the LAURA1 scheme is optimal and provides a benchmark for comparison to other possible simpler schemes.  In the following, two suboptimal power allocation strategies with reduced complexity are introduced. In Section \ref{SubSecLAURA1-CP}, a LAURA scheme with constant power source and relays is presented. In Section \ref{SubSec_LAURA1_CPR}, a LAURA scheme with adaptive power source transmission and constant power relays is presented.
\subsection{Constant Power Transmission (LAURA1-CP)}\label{SubSecLAURA1-CP}
A constant power solution to \eqref{Eq_Opt_Prob_General} is obtained by considering constant power transmission for the relays, i.e., $S_i^{(l)}=S$ and constant power transmission for the source, i.e., $S_{\text{s}}=S$ and removing step 4-2 in Algorithm 1. In this case, the SNR of  S-$\mathrm{UR}_i$ and $\gamma_{\text{eq}}$ are required at the S node as CSI to enforce the security and reliability constraints, respectively. The equivalent SNR, $\gamma_{\text{eq}}$, is estimated at the destination and fed back to the S node, or it can be calculated by S node knowing  CSI of S-D, S-$\mathrm{UR}_i$ and $\mathrm{UR}_i$-D.

The performance of LAURA1-CP is evaluated in Section \ref{Sec_Performance_Evaluation}. As elaborated below, for the case with one cooperating relay ($N_{\text{C}}=1$), the theoretical performance analysis is possible. Consider $i^*=\mathrm{arg}\max_i \gamma_{\text{eq}}^{(i)}$ obtained from step 5 of Algorithm 1. The event that relay $i^*$ satisfies the reliability constraint corresponds to the event that there is at least one relay that can satisfy this constraint. For LAURA1-CP with single cooperating relay, we have
\begin{equation}\label{Eq_Sec_And_reli_event_Constant_Pow_Mult_Relay}
A_n : \bigcap_{i=1}^{N_{\text{R}}}\gamma_{\text{s}i} \leq \mathrm{\Gamma}_n^{\text{r}} \,\cap\,
\bigcup_{i=1}^{N_{\text{R}}} \gamma_{\text{eq}}^{(i)} \geq \mathrm{\Gamma}_n^{\text{d}}.
\end{equation}
Then, the probability of selecting TM $m$ is
\begin{align}
P_m^{\text{CP}} \!\!=& \mathrm{Pr}\left(A_m
\cap \!\!\!\!\!\!\bigcap_{n=m+1}^N \!\!\!\!A_n^c\!\right)
=\mathrm{Pr}\left(\!\!\left\{\bigcap_{i=1}^{N_{\text{R}}}\gamma_{\text{s}i} \leq
\mathrm{\Gamma}_m^{\text{r}} \!\cap \bigcup_{i=1}^{N_{\text{R}}} \gamma_{\text{eq}}^{(i)}\geq
\mathrm{\Gamma}_{m}^{\text{d}}\!\right\}\!\!\cap\!\bigcap_{i=1}^{N_{\text{R}}} \gamma_{\text{eq}}^{(i)}<
\mathrm{\Gamma}_{m+1}^{\text{d}}\!\!\right)\nonumber\\
=&\mathrm{Pr}\left(\!\left\{\bigcap_{i=1}^{N_{\text{R}}}\gamma_{\text{s}i} \leq
\mathrm{\Gamma}_m^{\text{r}} \,\cap\, \gamma_{\text{eq}}^{(i)} <
\mathrm{\Gamma}_{m+1}^{\text{d}}\right\}\!\!\cap\bigcup_{i=1}^{N_{\text{R}}}
\gamma_{\text{eq}}^{(i)} \geq
\mathrm{\Gamma}_{m}^{\text{d}}\!\right)\nonumber\\ 
=&\mathrm{Pr}\left(\bigcap_{i=1}^{N_{\text{R}}}\gamma_{\text{s}i} \leq
\mathrm{\Gamma}_m^{\text{r}}\cap\gamma_{\text{eq}}^{(i)} \!<
\!\mathrm{\Gamma}_{m+1}^{\text{d}}\!\right)
-\mathrm{Pr}\left(\bigcap_{i=1}^{N_{\text{R}}}\gamma_{\text{s}i} \leq
\mathrm{\Gamma}_m^{\text{r}} \,\cap\, \gamma_{\text{eq}}^{(i)}<
\mathrm{\Gamma}_{m}^{\text{d}}\right).
\end{align}
Since different relay channels are independent, given $\gamma_{\text{sd}}$, $\gamma_{\text{eq}}^{(i)}$ is
independent of $\gamma_{\text{eq}}^{(j)}$ for $i\neq j$. Then
\begin{align}\label{Eq_Prob_of_Mode_General_MultRelay2}
P_m^{\text{CP}}\!\! =\mathbf{E}_{\gamma_{\text{sd}}}\Big\{\prod_{i=1}^{N_{\text{R}}}\mathrm{Pr}\left(\gamma_{\text{s}i}
\leq \mathrm{\Gamma}_m^{\text{r}} \,\cap\, \gamma_{\text{eq}}^{(i)}<
\mathrm{\Gamma}_{m+1}^{\text{d}}\vert\gamma_{\text{sd}}\right)
-\prod_{i=1}^{N_{\text{R}}}
\mathrm{Pr}\left(\gamma_{\text{s}i} \leq \mathrm{\Gamma}_m^{\text{r}} \,\cap\,
\gamma_{\text{eq}}^{(i)}< \mathrm{\Gamma}_{m}^{\text{d}}\vert\gamma_{\text{sd}}\right)\Big\}.
\end{align}
Using the upper bound for equivalent SNR in \eqref{Eq_MRC_UB}, a closed form expression for the TM probability is obtained as follows
\begin{align}\label{Eq_Prob_of_Mod_Apx}
P_m^{\text{CP}}\!\!=
&\mathbf{E}_{\gamma_{\text{sd}}}\Bigg\{\!\!\prod_{i=1}^{N_{\text{R}}}\!\!\left(\!\!1\!\!-\!e^{-\frac{\Gamma_m^{\text{r}}}{\bar \gamma_{\text{s}i}}}
\!\!-\!\!\left[e^{-\frac{\left[\Gamma_m^{\text{d}}-\gamma_{\text{sd}}\right]^+}{\bar\gamma_{\text{s}i}}}\!\!\!-e^{-\frac{\Gamma_m^{\text{r}}}{\bar\gamma_{\text{s}i}}}\right]^+ \!\!\!\!\!.\,e^{-\frac{\left[\Gamma_m^{\text{d}}-\gamma_{\text{sd}}\right]^+}{\bar\gamma_{i\text{d}}}}\!\!\right)\!\nonumber\\
-&\prod_{i=1}^{N_{\text{R}}}\left(\!\!1\!-\!e^{-\frac{\Gamma_m^{\text{r}}}{\bar \gamma_{\text{s}i}}}\!\! -\!\!\left[e^{-\frac{\left[\Gamma_{m+1}^{\text{d}}\!-\gamma_{\text{sd}}\right]^+}{\bar\gamma_{\text{s}i}}}\!-\!e^{-\frac{\Gamma_m^{\text{r}}}{\bar\gamma_{\text{s}i}}}\right]^+ \!\!\!\!\!\!.\,e^{-\frac{\left[\Gamma_{m+1}^{\text{d}}\!-\gamma_{\text{sd}}\right]^+}{\bar\gamma_{i\text{d}}}}\!\!\right)\!\!\!\Bigg\}\!.
\end{align}
Then, the average spectral efficiency is computed by substitution of \eqref{Eq_Prob_of_Mod_Apx} in \eqref{Eq_Spectral_Eff_General}. The authors also reported the above analysis in \cite{UntrustedRelayConferenceICT:Khodakarami}.

\subsection{Constant Power Relay Transmission (LAURA1-CPR)}\label{SubSec_LAURA1_CPR}
The LAURA1-CPR scheme involves adaptive power transmission for the source and constant power relay transmission. In this case, we set $S_i=S$ for $i\in \mathcal{M}_{\text{C}}$. The required CSI are S-$\mathrm{UR}_i$ and $\gamma_{\text{eq}}$. For $\gamma_{\text{eq}}$, we need to transmit from S node with the power  $S_{\text{s}}$, and wait for the estimation of $\gamma_{\text{eq}}$ at the destination or it can be calculated by S node knowing  CSI of S-D, S-$\mathrm{UR}_i$ and $\mathrm{UR}_i$-D. A modification of the Algorithm 1 yields the solution for constant power relay transmission, where, step 4-2 is replaced as follows
\begin{equation}\label{LAURA1-CP_Algorithm_Modify}
\text{Set  $S_i^{(l)}=S$ and $S_{\text{s}}^{(l)}=\min(S_{\text{tot}}-N_{\text{C}}S,S\dfrac{\Gamma_n^{\text{r}}}{\gamma_{\text{s}\tilde i}})$}.
\end{equation}
This substitutes solving the power allocation problem \eqref{Eq_Opt_Prob_Gamma_eq_General} with a simple power allocation that assumes power constraint for source and each relay separately.
\begin{remark0}\label{Remark_Separate_Power_Const}
In LAURA problem described in \eqref{Eq_Opt_Prob_General}, if we consider separate power constraints $S$ for source and each of the relays the total power constraint will still be $S_{\text{tot}}=(N_{\text{C}}+1)S$, the optimized solution will be LAURA1-CPR. This is due to the fact that the cooperating relays will need to transmit with their maximum allocated power to ensure maximized reliable transmission rate, and the source will be transmitting with its optimized adaptive power based on security and power constraints.
\end{remark0}
\section{LAURA: Relay Selection Strategies}\label{Sec_Relay_Selection_Strategies}
The solution presented for LAURA (problem \eqref{Eq_Opt_Prob_General}) in Algorithm 1 provides high performance but involves an exhaustive search over all $N_{\text{C}}$ member subsets of $\mathcal{M}_{\text{R}}$. Specifically, it examines optimized $\gamma_{\text{eq}}^{(l)}$ for the said relay subsets, which of course only a single one is finally used. In addition, the source requires $\gamma_{i\text{d}}$ for non-cooperating relays as CSI as well. Suboptimal solutions may have some advantages in practice, as their CSI requirements and complexity could be far less than the optimal one. In this section, we propose two efficient relay selection strategies for LAURA.
\subsection{Relay Selection Based on Source Relay CSI (LAURA2)}
An efficient solution with manageable CSI requirement may be constructed by taking a suboptimal relay selection approach that relies on $\gamma_{\text{s}i}$, $i\in \mathcal{M}_{\text{R}}$ and $\gamma_{\text{eq}}$ resulting from $\mathcal{M}_{\text{C}}$. Hence, there is no need to $\mathrm{UR}_i$-D  CSI feedback ($i\in \mathcal{M}_{\text{R}}$). To limit the complexity, we wish to avoid solving \eqref{Eq_Opt_Prob_Gamma_eq_General} to obtain $\gamma_{\text{eq}}$ \eqref{Eq_MRC_All_Relays} for all subsets of $\mathcal{M}_{\text{R}}$. To this end, the instantaneous source to relay SNR can be used as the relay selection criterion. Due to the security constraint, $\gamma_i$ is limited to $\frac{\Gamma_n^r\gamma_{\text{s}i}}{\gamma_{\text{s}\tilde i}}$. As a result, the subset of relays that provide a high $\gamma_{\text{eq}}$ may also be identified by selecting the relays with the highest $\gamma_{\text{s}i}$'s. Indeed, as we shall see in Section \ref{Sec_Performance_Evaluation}, this suboptimal and yet efficient relay selection strategy does not significantly degrade the performance of the optimal solution. Algorithm 2, gives this solution that is labeled as LAURA2.
\subsection{Constant Power Relay Transmission with Relay Selection Based on Source Relay CSI (LAURA2-CPR)}
The relay selection criterion according to SNR of source to relays can also be applied to the LAURA1 scheme with adaptive power source and constant power relays (Section \ref{SubSec_LAURA1_CPR}). This will further reduce the computational complexity and CSI requirements.  A modification of the Algorithm 2 yields the solution with adaptive power source and constant power relay transmission, where the step 4 is replaced with the following
\begin{equation*}
\text{Set  $S_i=S$ and $S_{\text{s}}=\min(S_{\text{tot}}-N_{\text{C}}S,S\dfrac{\Gamma_n^{\text{r}}}{\gamma_{\text{s}\tilde i}})$}.
\end{equation*}
\vspace{2mm}\\
\begin{tabular}{l}
\toprule[0.5mm]
Algorithm 2: LAURA with Modified Relay Selection\\
\midrule
1) Select $n=N$.\\
2) If $n=0$ then go to outage mode and exit.\\
3) Sort the set of available relays $\mathcal{M}_{\text{R}}$ according to their $\gamma_{\text{s}i}$ in descending order and  select $\mathcal{M}_{\text{C}}$
as the first $N_{\text{C}}$ \\ in the set.\\
4) Solve \eqref{Eq_Opt_Prob_Gamma_eq_General} and obtain $S_{\text{s}}$ and $S_i$ for $i\in \mathcal{M}_{\text{C}}$.\\ 
5) Calculate $\gamma_{\text{eq}}$.\\
6) If $\mathrm{C2}$ in \eqref{Eq_Opt_Prob_General} is satisfied, set $\mathrm{TM}=n$ and exit; else $n=n-1$ and go to step $2$.
\vspace{1mm}\\
\bottomrule[0.5mm]
\end{tabular}
\vspace{2mm}

\subsection{Constant Power Relay Transmission with Relay Selection Based on Source Relay Channel Statistics (LAURA3-CPR)}\label{Section_LAURA3_CPR}
The instantaneous relay selection criterion according to SNR of source to relays involves high speed (per frame) switching of relays and hence a rather sizable network control overhead. A relay selection criterion according to the statistics of source to relay channels can help mitigate this problem. This will further reduce the computational complexity and CSI requirements since the selected relays are fixed as long as the average SNRs of source to relays remain unchanged.  A modification of the Algorithm 2 yields the solution for adaptive power source transmission and constant power relay transmission in which the step 3 is replaced as follows
\vspace{2mm}\\
Sort the set of available relays $\mathcal{M}_{\text{R}}$ according to their $\bar\gamma_{\text{s}i}$ in descending order and select $\mathcal{M}_{\text{C}}$
as the first $N_{\text{C}}$ in the set.\\
\vspace{2mm}

The selection of relays according to the average channel conditions also makes the theoretical performance analysis of the system possible. In the following, we present a performance analysis of LAURA3-CPR.
The event $A_n$ in this case using the upper bound for equivalent SNR is expressed by
\begin{align}\label{Eq_Prob_of_Mode_Can_be sel_Mult_Coop_Adaptive_power}
&\mathrm{Pr} (A_n)=\mathrm{Pr} \left(\gamma_{\text{eq,u}}\geq\Gamma_n^{\text{d}}\,|\, S_{\text{s}} =  \tilde S_{\text{s},n}(\gamma_{\text{s}\tilde i})\right),
\end{align}
where $\tilde S_{\text{s},n}(\gamma_{\text{s}\tilde i})=\min(S_{\text{tot}}-N_{\text{C}}S,S\frac{\Gamma_n^{\text{r}}}{\gamma_{\text{s}\tilde i}})$. In order to calculate the probability of event $A_n$, we first derive the moment generating function (MGF) of $\gamma_{\text{eq,u}}$  given
$\gamma_{\text{s}\tilde i}$ and $S_{\text{s}} =\tilde S_{\text{s},n}(\gamma_{\text{s}\tilde i})$. Since $\frac{S_{\text{s}}}{S}\gamma_{\text{sd}}$ and $\{\gamma_{i,\text{u}}\}$ are all independent of each other given $\gamma_{\text{s}\tilde i}$, the desired MGF knowing $S_{\text{s}} = \tilde S_{\text{s},n}(\gamma_{\text{s}\tilde i})$ is given by
\begin{equation}
M_{\gamma_{\text{eq,u}}|\gamma_{\text{s}\tilde i}}(s|x) = M_{\frac{S_{\text{s}}}{S}\gamma_{\text{sd}}|\gamma_{\text{s}\tilde i}}(s|x)\prod_{i \in \mathcal{M}_{\text{C}}}M_{\gamma_{i,\text{u}}|\gamma_{\text{s}\tilde i}}(s|x),
\end{equation}
where $M_{\frac{S_{\text{s}}}{S}\gamma_{\text{sd}}|\gamma_{\text{s}\tilde i}}(s|x)$ and $M_{\gamma_{i,\text{u}}|\gamma_{\text{s}\tilde i}}(s|x)$ are respectively the MGF of $\frac{S_{\text{s}}}{S}\gamma_{\text{sd}}$ and $\gamma_{i,\text{u}}$ given $\gamma_{\text{s}\tilde i}$ and $S_{\text{s}} = \tilde S_{\text{s},n}(\gamma_{\text{s}\tilde i})$.
Using the definition of the MGF as $M_X(s) = \mathbf{E}\left(e^{-sX}\right)$, it can
be easily shown that
\begin{equation}
M_{\frac{S_{\text{s}}}{S}\gamma_{\text{sd}}|\gamma_{\text{s}\tilde i}}(s|x)=\frac{1}{1+s\cdot\, \frac{\tilde S_{\text{s},n}(x)}{S}}.
\end{equation}
In order to calculate $M_{\gamma_{i,\text{u}}|\gamma_{\text{s}\tilde i}}(s|x)$ we first calculate cumulative distribution function (CDF) of  $\gamma_{i,\text{u}}$ as follows
\begin{align}
F_{\gamma_{i,\text{u}}|\gamma_{\text{s}\tilde i}}(z)  = \mathrm{Pr}(\gamma_{i,\text{u}}\leq z|\gamma_{\text{s}\tilde i} = x)
 =1-\mathrm{Pr}(\frac{S_{\text{s}}}{S}\gamma_{\text{s}i}\geq z|\gamma_{\text{s}\tilde i} = x)\mathrm{Pr}(\gamma_{i\text{d}}\geq z),
\end{align} 
where $\mathrm{Pr}(\gamma_{i\text{d}}>z)=e^{\frac{-z}{\bar\gamma_{i\text{d}}}}$. Using the conditional PDF of $f_{\gamma_{\text{s}i}|\gamma_{\text{s}\tilde i}}(y|x)$ according to Appendix \ref{Joint_PDF_Of_G_si_and_max_G_si}, we have
\begin{align}
&\mathrm{Pr}(\frac{S_{\text{s}}}{S}\gamma_{\text{s}i}>z|\gamma_{\text{s}\tilde i} = x) = \int_{\frac{z}{\tilde S_{\text{s},n}(x)/S}}^{\infty} f_{\gamma_{\text{s}i}|\gamma_{\text{s}\tilde i}}(y|x) \ud y \nonumber\\ 
&= \int_{\frac{z}{\tilde S_{\text{s},n}(x)/S}}^{\infty} \left[\frac{B_i(x)}{C(x)}\frac{1}{\bar\gamma_{\text{s}i}}e^{-\frac{y}{\bar\gamma_{\text{s}i}}}\mathcal{U}(x-y) +\frac{D_i(x)}{C(x)}\delta(y-x)\right]\ud y\nonumber\\
&=\! \left[\frac{B_i(x)}{C(x)}\left(\!e^{-\frac{S\,z}{\tilde S_{\text{s},n}(x)\bar\gamma_{\text{s}i}}}\!-\!e^{-\frac{x}{\bar\gamma_{\text{s}i}}}\!\right)\! +\!\frac{D_i(x)}{C(x)}\right]\!\mathcal{U}(x-\frac{S\,z}{\tilde S_{\text{s},n}(x)}), 
\end{align}
where $\mathcal{U}(\cdot)$ and $\delta(\cdot)$ are unit step and unit impulse functions, respectively, and  $B_i(x)$, $C(x)$ and $D_i(x)$ are defined in Appendix \ref{Joint_PDF_Of_G_si_and_max_G_si}. Then,
\begin{align}
M_{\gamma_{i,\text{u}}|\gamma_{\text{s}\tilde i}}(s|x) &= s \int_0^{\infty} F_{\gamma_{i,\text{u}}|\gamma_{\text{s}\tilde i}}(z|x) e^{-s z} \ud z\nonumber\\
&= 1-\frac{B_i(x)}{C(x)}\frac{s}{\frac{S}{\tilde S_{\text{s},n}(x) \bar\gamma_{\text{s}i}}+\frac{1}{\bar\gamma_{i\text{d}}}+s}\cdot
\left(1-e^{-x \frac{\tilde S_{\text{s},n}(x)}{S}\left(\frac{S}{\tilde S_{\text{s},n}(x) \bar\gamma_{\text{s}i}}+\frac{1}{\bar\gamma_{i\text{d}}}+s\right)}\right)\nonumber\\
&\hspace{8mm}+\left(\frac{B_i(x)}{C(x)}e^{\frac{-x}{\bar\gamma_{\text{s}i}}}-\frac{D_i(x)}{C(x)}\right)\cdot
\frac{s}{\frac{1}{\bar\gamma_{i\text{d}}}+s}\left(1-e^{-x\frac{\tilde S_{\text{s},n}(x)}{S}\left(\frac{1}{\bar\gamma_{i\text{d}}}+s\right)}\right).
\end{align}
Finally,
\begin{align}\label{Eq_Prob_of_Mode_Can_be sel_Mult_Coop_Adaptive_power_Final}
\mathrm{Pr} (A_n)=\mathrm{Pr} \left(\gamma_{\text{eq}}\geq\Gamma_n^{\text{d}}\right) = 1- \mathbf{E}_{x}\left\{\mathrm{Pr} \left(\gamma_{\text{eq}}\leq\Gamma_n^{\text{d}}\right)|\gamma_{\text{s}\tilde i} = x\right\}
= 1- \mathbf{E}_{x}\left\{\mathfrak{L}^{-1}\left\{\frac{M_{\gamma_{\text{eq}}|\gamma_{\text{s}\tilde i}}(s|x)}{s}\right\}_{\gamma_{\text{eq}} = \Gamma_n^{\text{d}}}\right\},
\end{align}
where $\mathfrak{L}^{-1} $ denotes inverse Laplace transform with respect to $\gamma_{\text{eq,u}}$ that is simply computed through symbolic evaluation with MATLAB for every $\mathcal{M}_{\text{R}}$ and $\mathcal{M}_{\text{C}}$. Then, the expectation with respect to $\gamma_{\text{s}\tilde i}$ is computed through numerical integration.

Following \eqref{Eq_Prob_of_Mode_Ada_Pow_Mult_Relay_SC}, the probability of TM $m$ for LAURA3-CPR is expressed as 
\begin{align}\label{Eq_Prob_of_Mode_Adaptive_Power_MC}
P_m^{\text{LAURA3-CPR}}=\mathrm{Pr}\left(\bigcup_{n=m}^N
A_n\right)-\mathrm{Pr}\left(\bigcup_{n=m+1}^N A_n\right)
=\mathbf{E}\left\{\mathbf{I}\left(A_{\tilde n_m}\right)-\mathbf{I}\left(A_{\tilde n_{m+1}}\right)\right\},
\end{align}
where
\begin{align}\label{n_tilde_MCP_MC}
\tilde {n}_m(\gamma_{\text{eq,u}},\gamma_{\text{s}\tilde i})
= \mathrm{arg}\min_{n\in \{m,...,N\}}\left(\gamma_{\text{eq,u}}\geq\Gamma_n^{\text{d}}\,|\, S_{\text{s}} = \tilde S_{\text{s},n}(\gamma_{s \tilde i})\right).
\end{align}
In order to facilitate the analysis we consider an approximation like the one in \eqref{n_tilde_Approximation} as
\begin{equation}\label{n_tilde_Approximation_MC}
\tilde {n}_m (\gamma_{\text{eq,u}},\gamma_{\text{s}\tilde i})\approx m.
\end{equation}
And finally, the probability of TM $m$ is
\begin{align}\label{Eq_Prob_of_Mode_Adaptive_Power_MC2}
P_m^{\text{LAURA3-CPR}}=\mathrm{Pr}\left(A_{m}\right)-\mathrm{Pr}\left(A_{m+1}\right).
\end{align}
Calculating the probability of TM based on \eqref{Eq_Prob_of_Mode_Adaptive_Power_MC2} and \eqref{Eq_Prob_of_Mode_Can_be sel_Mult_Coop_Adaptive_power_Final}, enables the computation of average spectral efficiency according to \eqref{Eq_Spectral_Eff_General}.
\begin{remark0}\label{Remark_N_c_N_R}
Since for $N_{\text{C}}=N_{\text{R}}$ no relay selection is employed, the LAURA performance is independent of the relay selection strategy. Hence, the presented performance evaluation of LAURA3-CPR in \eqref{Eq_Prob_of_Mode_Adaptive_Power_MC2} (and \eqref{Eq_Prob_of_Mode_Can_be sel_Mult_Coop_Adaptive_power_Final}) also applies to LAURA1-CPR and LAURA2-CPR schemes.
\end{remark0}
\section{Performance Evaluation}\label{Sec_Performance_Evaluation}
In this section, the performance of the proposed LAURA1, LAURA2 and LAURA3 schemes of different relay selection strategies in conjunction with different power control mechanisms of adaptive power, CP and CPR are evaluated. Both analytical and numerical results are presented and the effects of different parameters on the performance are investigated. 
\subsection{Experiment Setup}
Figure \ref{Network} illustrates the topology of the network under consideration with $N_{\text{R}}$ available relays. We consider average SNR of each channel proportional to $\frac{1}{\ell^{\alpha}}$, due to path loss, where $\ell$ denotes the distance between the two parties ($\alpha$ is set to $4$ in this paper). Without loss of generality we consider the distance between S and D nodes normalized to $1$. Relays are all located on a line perpendicular to the line connecting S and D nodes each distanced $0.1$ apart as depicted in Figure \ref{Network}. As shown in Fig. \ref{Network}, $\mathrm{UR}_1$ node is fixed at the distance $\ell_{\text{s}1}$ on the line connecting S and D nodes. In all figures, $\mathrm{BER}_{\text{tgt}}^{\text{r}}$ and $\mathrm{BER}_{\text{tgt}}^{\text{d}}$ are set to $0.1$ and $10^{-6}$, respectively. The sum power constraint $S_{\text
 {tot}}$ is set to $(N_{\text{R}}+1)S$ in all cases.

\begin{figure}[!t]
\centering\includegraphics[width=3.5in]{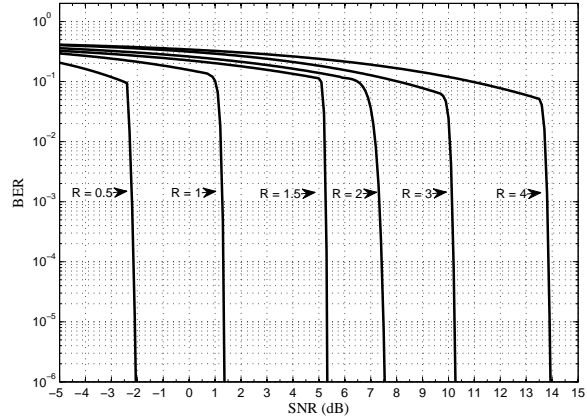}
\caption{BER curves for six transmission modes and their corresponding rates}
\label{AMCModesBERCurves}
\end{figure}

The TMs used with the proposed LAURA schemes could be in fact any set of possible channel coding and modulation pairs. The suggestion is to use sharp channel codes to obtain sharp BER curves for TMs, which provide a low security gap and an acceptable reliability performance (for the design parameter ranges of interest). The set of LDPC codes in DVB-S2 standard \cite{DVBS2:Standard} offer this characteristic. These channel codes in conjunction with different modulation schemes yield the BER curves depicted in Fig. \ref{AMCModesBERCurves}. The TMs specifications and fitting parameters used for numerical results are presented in Table \ref{TransmissionModesTable}. 
\linespread{1.5}
\begin{table}[!t]
\centering 
\caption{Transmission Modes for AMC Scheme and Their
Corresponding Fitting Parameters}\label{TransmissionModesTable}
\begin{tabular}{|c|c|c|c|c|c|c|}
\hline
\textbf{Mode}&\textbf{1}&\textbf{2}&\textbf{3}&\textbf{4}&\textbf{5}& \textbf{6} \\
\hline
\vspace{-4mm}  & & & & & & \\ \hline
\hspace{-2mm}modulation\hspace{-2mm}              & \hspace{-2mm}BPSK \hspace{-2mm} &\hspace{-2mm} 4-QAM \hspace{-2mm}&  \hspace{-2mm}8-QAM \hspace{-2mm} &\hspace{-2mm} 8-QAM & \hspace{-2mm} 16-QAM\hspace{-2mm} &\hspace{-2mm} 32-QAM \hspace{-2mm}\\
\hspace{-1mm}coding rate \hspace{-2mm}            & 1/2   & 1/2   & 1/2     & 2/3   & 3/4     & 4/5 \\
\hspace{-3mm}rate:$R_n$ \hspace{-2mm} & 1/2   & 1     & 3/2     & 2     & 3       & 4 \\
\hline
\vspace{-4mm}  & & & & & & \\ \hline
\hspace{-2mm}$p_n$      \hspace{-2mm}             & 2.97 & 1.17 & 0.80     & 0.65 & 0.46   & 0.38 \\
\hspace{-2mm}$q_n$   \hspace{-2mm}                & 1.05  & 0.68 & 0.54   & 0.59 & 0.67   & 0.58 \\
\hspace{-2mm}$a_n$   \hspace{-2mm}                & 1.55 & 0.13 & 0.14   & 0.12& 0.06   & 0.07 \\
\hspace{-2mm}$b_n$   \hspace{-2mm}                & 0.62 & 1.43 & 3.48   & 6.19 & 10.37   & 24.09 \\
\hspace{-2mm}$c_n$    \hspace{-2mm}               & 41.9  & 34.90 & 20.03   & 3.52 & 5.13    & 2.21 \\
\hspace{-2mm}$k_n$   \hspace{-2mm}               & 16.81 & 104.70 & 48.83   & 75.99& 6.77   & 6.50 \\
\hline
\end{tabular}
\end{table}
\linespread{1.9}
\subsection{Numerical Results}
\begin{figure}[!t]
\centering
\includegraphics[width=3.5in]{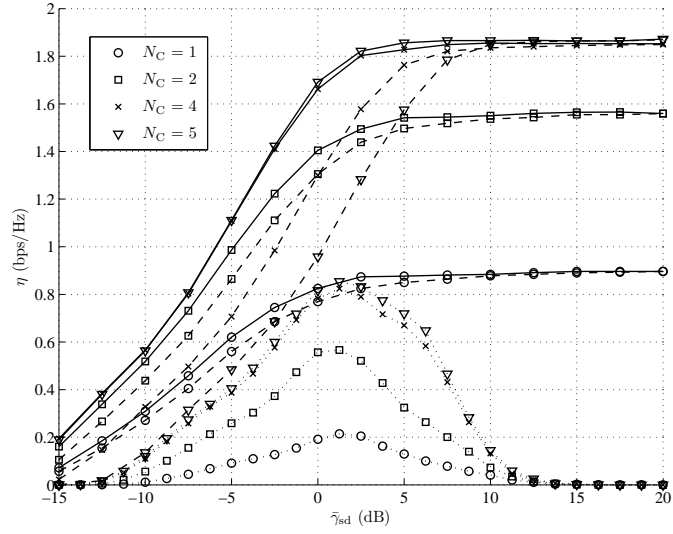}
\caption{Average spectral efficiency of different transmission power strategies with optimal relay selection for $N_{\text{R}}=5$ and $\ell_{\text{s}1} = 0.9$. Solid, dashed and dotted lines are for LAURA1, LAURA1-CPR and LAURA1-CP, respectively.}
\label{SE_CPR_APR_CP_Compare}
\end{figure}

Figure \ref{SE_CPR_APR_CP_Compare} depicts the end to end average spectral efficiency of LAURA1, LAURA1-CPR and LAURA1-CP schemes illustrating the effect of type of power control. The horizontal axis indicates $\bar\gamma_{\text{sd}}$. It is observed that for LAURA1-CP, increasing average SNR of S-D link  beyond a specific value leads to a reduced average spectral efficiency. This is due to the limitation in number of TMs. If there is a TM that could still satisfy the security constraint in high SNRs, then the declination occurs later at a higher average SNR of S-D channel. Comparing the performance of LAURA1 and LAURA1-CP highlights the advantage of source power adaptation that significantly reduces the outage events for enforcing the security constraint.

The performance comparison of LAURA1 and LAURA1-CPR suggests that for low to medium SNR regimes relay power adaptation improves the performance of the system. This is more significant with larger $N_{\text{C}}$'s. One sees from the performance of LAURA1 and LAURA1-CP that the larger the number of cooperating relays, the better the performance we achieve through cooperation. However, there is no significant advantage in using more than four relays for $N_{\text{R}}=5$. This observation is in contrast to the understanding in traditional (non-secure) amplify and forward relaying that the best relay selection ($N_{\text{C}}=1$) achieves almost all of the spectral efficiency performance gain \cite{AsimpleCooperativePathSelection:Reed}. 

Comparing LAURA1-CPR for $N_{\text{C}}=4$ and $5$ in low SNR regimes reveals that with constant power relays and sum power constraint, increasing the number of cooperating relays beyond a limit may decrease the performance. The reason is that in this case, the available power to the S-node (and hence the SNRs of source to relays) becomes limited. 

\begin{figure}[!t]
\centering
\includegraphics[width=3.5in]{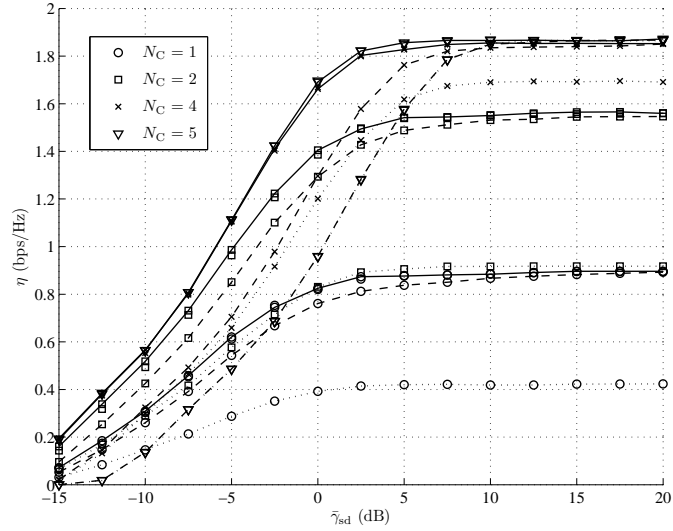}
\caption{Average spectral efficiency of different relay selection schemes for $N_{\text{R}}=5$ and $\ell_{\text{s}1} = 0.9$. Solid, dashed and dotted lines are for LAURA1, LAURA2-CPR and LAURA3-CPR, respectively. No line is used to connect the markers associated with LAURA2, but they closely follow those of LAURA1. Note that the performance of LAURA2-CPR and LAURA3-CPR for $N_{\text{C}}=5$ coincide.}
\label{SE_RelaySelect_Compare}
\end{figure}

Figure \ref{SE_RelaySelect_Compare} illustrates the effect of relay selection strategies on the average spectral efficiency performance of LAURA schemes. The main observation is that the performance of relay selection according to source-relay channel SNRs coincides with that of optimal relay selection. This is evident both by performance comparison of LAURA1 with LAURA2 and LAURA1-CPR with LAURA2-CPR. It is also observed that the performance of LAURA3-CPR with relay selection according to average source-channel SNRs approaches that of LAURA2-CPR utilizing instantaneous channel SNRs only when the number of relays increases to $N_{\text{C}}=5$. Nevertheless, the relay selection according to average channel statistics requires slower relay switching.

\begin{figure}[!t]‎
‎\centering‎
‎\includegraphics[width=3.5in]{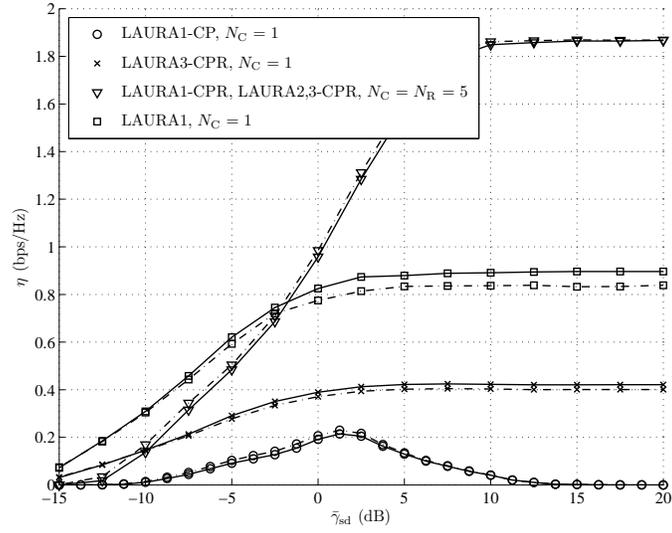}‎
\caption{Comparison of analytical and simulation results for average spectral efficiency of different LAURA schemes for $N_{\text{R}}=5$, $\ell_{\text{s}1} = 0.9$. Solid and dashed lines depict the simulation and analytical results, respectively‎.}‎
‎\label{SE_Sim_Theory_Compare}‎
‎\end{figure}

Figure ‎\ref{SE_Sim_Theory_Compare} depicts and compares the simulation and approximate analytical results for certain LAURA schemes. As discussed in Remark \ref{Remark_N_c_N_R}, for $N_{\text{C}}=N_{\text{R}}$, since no relay selection is employed, the theoretical calculation of spectral efficiency proposed for \ref{Section_LAURA3_CPR} applies to all the schemes with constant power relay transmission. The results verify the accuracy of the proposed analytical performance evaluations.

\begin{figure}[!t]
\centering
\includegraphics[width=3.5in]{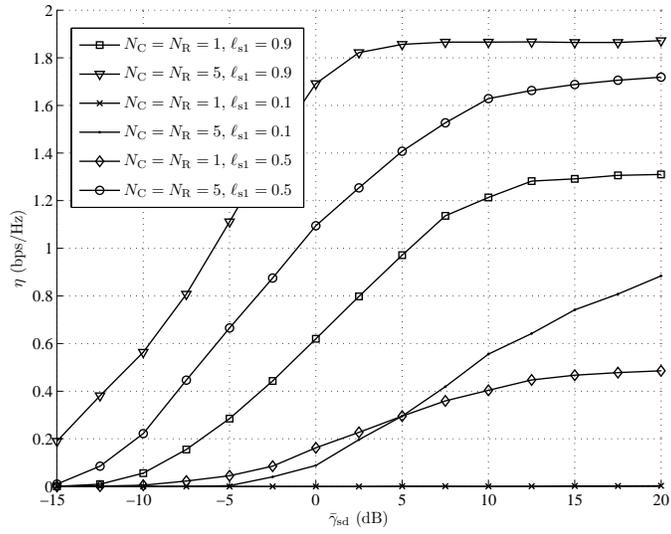}
\caption{Average spectral efficiency of LAURA1 scheme for different relay placements.}
\label{SE_LAURA_Relay_Placement}
\end{figure}

Figure \ref{SE_LAURA_Relay_Placement} demonstrates the effect of relays positions on the performance of the proposed LAURA scheme. It is evident that when the relays are positioned closer to the D node, LAURA provides a better average spectral efficiency performance. In fact, in this setting the security constraint is satisfied more easily. Our experiments reveal that in an amplify and forward cooperative communication system, in the SNR ranges of interest for cooperation (low S-D SNR regimes), when a sufficiently large number of available relays are utilized, provisioning the security constraint in LAURA only imposes a negligible spectral efficiency performance penalty. In fact, for larger number of cooperating relays, the performance penalty is small even for high SNR regime, e.g., less than $7\%$ for $N_{\text{C}}=N_{\text{R}}=5$.

\section{Conclusions}\label{ConclusionSection}
In this paper, a link adaptation and untrusted relay assignment framework for cooperative communications with physical layer security is proposed. The design problem is set up for highly spectrally efficient communications with reliability for the destination and security in the presence of untrusted relays. The security constraint is imposed by ensuring that the relays cannot decode useful information from the signal they relay. The optimal solution to the design optimization problem is presented while strategies for mitigating practical challenges are also proposed. This involves several relay selection strategies and power adaptation solutions. Performance of these approaches is analyzed theoretically, in certain cases, and rigorously through simulations. The effect of different design strategies and parameters, including relay selection, power control at source and/or relays, relays positions, and number of cooperating relays are investigated. 

The future research of interest in this direction includes tackling potential eavesdroppers in the network or considering the malicious behavior of relays without a service level trust. Another orientation of research interest is to design particular channel codes with low security gap and reliability performance within this framework and in line with the works reported in \cite{LDPCforAWGN:Mclaughlin}.

\appendices
\section{}\label{Joint_PDF_Of_G_si_and_max_G_si}
In this section we derive the joint and conditional PDF of $\gamma_{\text{s}\tilde i}$ and $\gamma_{\text{s}i}$. The joint cumulative distribution function of  these two variables is given by
\begin{align}
F_{\gamma_{\text{s}\tilde i},\gamma_{\text{s}i}}(x,y) = \mathrm{Pr}(\max_j \gamma_{\text{s}j} <x , \gamma_{\text{s}i}<y) =
\prod_{\substack{j=1\\ j\neq i}}^{N_{\text{R}}}\left(1-e^{-\frac{x}{\bar\gamma_{\text{s}j}}}\right)
 \left[\left(1-e^{-\frac{x}{\bar\gamma_{\text{s}i}}}\right)\mathcal{U}(y-x) +\left(1-e^{-\frac{y}{\bar\gamma_{\text{s}i}}}\right)\mathcal{U}(x-y) \right].\nonumber
\end{align}
Then, the joint probability density function of $\gamma_{\text{s}\tilde i}$ and $\gamma_{\text{s}i}$ is
\begin{align}\label{Eq_Joint_PDF_Max_Gsi_Gsi}
 f_{\gamma_{\text{s}\tilde i},\gamma_{\text{s}i}}(x,y)=& \frac{\partial^2 F_{\gamma_{\text{s}\tilde i},\gamma_{\text{s}i}}(x,y) }{\partial x\partial y}\nonumber\\
=&\sum_{\substack{l\neq i\\l=1}}^{N_{\text{R}}}\frac{1}{\bar\gamma_{\text{s}l}}e^{-\frac{x}{\bar\gamma_{\text{s}l}}}\prod_{\substack{j\neq l\\j\neq i}}\left(1-e^{-\frac{x}{\bar\gamma_{\text{s}j}}}\right)\left[\frac{1}{\bar\gamma_{\text{s}i}}e^{-\frac{y}{\bar\gamma_{\text{s}i}}}\mathcal{U}(x-y) \right]
+\prod_{\substack{j\neq i\\j=1}}^{N_{\text{R}}}\left(1-e^{-\frac{x}{\bar\gamma_{\text{s}j}}}\right)\frac{1}{\bar\gamma_{\text{s}i}}e^{-\frac{x}{\bar\gamma_{\text{s}i}}}\delta(y-x)\nonumber\\
\triangleq & B_i(x)\left[\frac{1}{\bar\gamma_{\text{s}i}}e^{-\frac{y}{\bar\gamma_{\text{s}i}}}\mathcal{U}(x-y) \right]+D_i(x)\delta(y-x) ,
\end{align}
and the PDF of $\gamma_{\text{s}\tilde i}$ is simply
\begin{equation}
f_{\gamma_{\text{s}\tilde i}}(x)=\sum_{l=1}^{N_{\text{R}}}\frac{1}{\bar\gamma_{\text{s}l}}e^{-\frac{x}{\bar\gamma_{\text{s}l}}}\prod_{j\neq l}\left(1-e^{-\frac{x}{\bar\gamma_{\text{s}j}}}\right)\triangleq C(x) .
\end{equation}
The PDF of $\gamma_{\text{s}i}$ given $\gamma_{\text{s}\tilde i}$ is then
\begin{align}\label{Eq_Conditional_PDF_Max_Gsi_Gsi}
f_{\gamma_{\text{s}i}|\gamma_{\text{s}\tilde i}}&(y|x)=  \frac{f_{\gamma_{\text{s}\tilde i},\gamma_{\text{s}i}}(x,y)}{ f_{\gamma_{\text{s}\tilde i}}(x)}
= \frac{B_i(x)}{C(x)}\left[\frac{1}{\bar\gamma_{\text{s}i}}e^{-\frac{y}{\bar\gamma_{\text{s}i}}}\mathcal{U}(x-y) \right]+\frac{D_i(x)}{C(x)}\delta(y-x).
\end{align}

\bibliographystyle{IEEEtran}
\bibliography{IEEEabrv,Journal}
\newpage

‎

\end{document}